\documentclass[sigconf,10pt]{acmart}
\usepackage{listings}
\usepackage{color}
\usepackage{epsfig,makecell}                                      
\hypersetup{citecolor=blue,linkcolor=blue}                        
\usepackage{thmtools}
\usepackage{thmbox}
\usepackage[toc,page]{appendix}
\usepackage{endnotes,microtype,xspace,fancyvrb,multirow} 
\usepackage{epigraph} 

\usepackage{booktabs}      
\usepackage{tabularx}
\usepackage{array,underscore,relsize}                             
\usepackage[T1]{fontenc}                                          
\usepackage{microtype}     
\usepackage{times}                                                
\usepackage{enumitem}                                             
\usepackage[labelfont=bf,font=normal,skip=1pt]{caption}           
\usepackage[export]{adjustbox}                                    
\usepackage{breakurl}                                             
\usepackage{pifont}                                               
\usepackage{tablefootnote}                                        
\usepackage{bbding}                                        
\usepackage[linesnumbered,vlined,boxed,commentsnumbered]{algorithm2e}
\DontPrintSemicolon
\newcommand{\ra}[1]{\renewcommand{\arraystretch}{#1}}
\usepackage{algpseudocode}
\usepackage[normalem]{ulem}
\usepackage[hang,flushmargin]{footmisc}
\usepackage{textcomp}
\usepackage{graphicx}
\usepackage[scaled]{beramono}
\usepackage{subfig}
\usepackage{float}
\usepackage{listings}
\usepackage{courier}
\usepackage{color}
\usepackage{colortbl}
\usepackage{stfloats}

\usepackage{stfloats}

\usepackage[compact]{titlesec}
\titleformat*{\section}{\Large\bfseries}
\titleformat*{\subsection}{\normalsize\bfseries}
\titleformat*{\subsubsection}{\normalsize\bfseries}
\titlespacing{\section}{0pt}{3ex}{1.5ex}
\titlespacing{\subsection}{0pt}{2ex}{1ex}
\titlelabel{\thetitle\hspace{-1mm}\quad}

\hypersetup{
  colorlinks,
  linkcolor={red!50!black},
  citecolor={blue!50!black},
  urlcolor={blue!80!black}
}

\newcommand{\fig}[1]{Figure{~\ref{#1}}}

\newcommand{\one}{\texttt{\uppercase\expandafter{\romannumeral1}}}
\newcommand{\two}{\texttt{\uppercase\expandafter{\romannumeral2}}}

\newcommand{\XD}[1]{\textcolor{blue}{XD: #1}}

\newcommand{\TODO}[1]{\textcolor{red}{TODO: #1}}

\newcommand{\revise}[1]{{\textcolor{red}{#1}}}

\newcommand{\stitle}[1]{\vspace{1.2ex}\noindent{\bf #1}\,}
\newcommand{\nospacestitle}[1]{\noindent{\bf #1}\,}
\newcommand{\etitle}[1]{\vspace{0.8ex}\noindent{\em\uline{#1}}}

\newcommand{\sys}{\textsc{KunServe}}

\newcommand{\sota}{state-of-the-art}
\newcommand{\eg}{e.g.,~}
\newcommand{\ie}{i.e.,~}

\definecolor{list_gray}{rgb}{0.5,0.5,0.5}
\definecolor{list_blue}{rgb}{0.13,0.13,1}
\definecolor{list_green}{rgb}{0,0.5,0}
\definecolor{list_mauve}{rgb}{0.58,0,0.82}
\usepackage{tikz}

\makeatletter
\renewcommand\@titlefont{\huge\normalfont\bfseries} 
\makeatother

\usepackage{balance}
\acmYear{2026}\copyrightyear{2026}
\acmConference[EUROSYS '26]{European Conference on Computer Systems}{April 27--30, 2026}{Edinburgh, Scotland Uk}
\acmBooktitle{European Conference on Computer Systems (EUROSYS '26), April 27--30, 2026, Edinburgh, Scotland Uk}
\acmDOI{10.1145/3767295.3769348}
\acmISBN{979-8-4007-2212-7/26/04}

\begin{document}

\title[Parameter-centric Memory Management for LLM Serving]{{\sys}: Parameter-centric Memory Management for Efficient Memory Overloading Handling in LLM Serving}

\author{\Large
  \vspace{-1mm}
  Rongxin Cheng \ \ Yuxin Lai$^{\dagger}$ \ \
  Xingda Wei{\Envelope} \ \ Rong Chen \ \ Haibo Chen \\ \vspace{2mm}
  {{Institute of Parallel and Distributed Systems, Shanghai Jiao Tong University}}
  \vspace{2mm}
}

\renewcommand{\shortauthors}{R. Cheng, Y. Lai, X. Wei, R. Chen, and H. Chen}

\date{}

\begin{abstract}

\noindent
\noindent
Serving LLMs with a cluster of GPUs is common nowadays,
where the serving system must meet strict latency SLOs required by applications.
However, the stateful nature of LLM serving requires maintaining huge states (\ie{KVCache})
in limited GPU memory.
Under spikes in real-world workloads,
GPU memory can be easily overloaded,
leading to orders of magnitude higher response latency due to queuing 
introduced by waiting for KVCache to be reclaimed. 
Prior KVCache-centric approaches handle overloading by
dropping, migrating, or swapping KVCache.
These methods fail to release sufficient memory quickly with 
requests still queued. 

This paper proposes the first parameter-centric approach to handling overloading
by selectively dropping replicated parameters to instantly free memory for requests,
based on an unnoticed observation that model parameters are commonly replicated
across GPUs for serving LLMs.
With additional memory, all requests can be served with a larger batch without queuing.
To make the parameter-centric approach correct and efficient,
we cooperatively execute requests on GPUs
with a complete copy of parameters using pipeline parallelism,
and derive an appropriate drop plan without unnecessary cooperation.
We also design techniques to minimize the performance overhead due
to pipeline parallelism with the execution patterns of requests under drop.
Evaluations show that {\sys} reduces the tail TTFT of
requests under overloading by up to {72.2\,$\times$} compared to the state-of-the-art
systems including Llumnix, vLLM and InferCept.

\end{abstract}


\keywords{LLM Serving; Cloud computing; Parameter-centric memory management\vspace{1mm}} 

\maketitle

\def\thefootnote{$\dagger$}\footnotetext{Work done while Yuxin was an intern at Institute of 
Parallel and Distributed Systems, Shanghai Jiao Tong University. 
Yuxin was affiliated with Huazhong University of Science and Technology.}
\def\thefootnote{{\Envelope}}\footnotetext{Xingda Wei is the corresponding author (\url{wxdwfc@sjtu.edu.cn}).}

\renewcommand{\thefootnote}{\arabic{footnote}}


\section{Introduction}
\label{sec:intro}

\noindent
Transformer-based large language models (LLMs) are reshaping the computing industry,
which
generate output in a token-by-token streaming fashion with auto-regressive inference.
The tokens are used by downstream tasks such as chatbots~\cite{chatgpt},
copilots~\cite{copilot},
and interactive agents~\cite{DBLP:conf/iclr/FurutaLNMFGG24}.
Such tasks require human interaction,
so serving LLMs has tight latency requirements,
\eg{less than 1 second~\cite{latency-study,distserve}}.
The smaller, the better~\cite{aws-latency}.
Specifically, both the time to generate the first token (TTFT)
and the time between subsequent tokens (TPOT) are important metrics.

A key feature of LLM inference is that the
computation is \emph{stateful}:
before generating the final token, the intermediate results of previously generated tokens (termed \emph{KVCache})
are kept in the scarce GPU memory (HBM) to accelerate future token generation.
Such a stateful generation introduces a key issue:
the serving latency could spike
(up to {239}\,$\times$ in BurstGPT~\cite{burstgpt},
see \textsection{\ref{sec:bg-challenges}} and others in \textsection{\ref{sec:eval}})
when the stored KVCache exhausts the precious HBM.
Such overloading is common under real-world request bursts~\cite{splitwise,DBLP:conf/osdi/FuXHBUPM24}
since the KVCache is proportional to the number of requests
processed (or to be processed).
Such overloading significantly impacts latency,
because requests must wait for GPUs to free up sufficient memory for processing.
Unfortunately, it could take seconds for LLMs to generate the final token so as to release memory due to the long and unpredictable
token generation process.

State-of-the-art approaches adjust KVCache stored in GPU memory to handle 
overloading~\cite{vllm,DBLP:journals/corr/abs-2404-09526,DBLP:journals/corr/abs-2407-00079,llumnix}.
When a GPU lacks sufficient HBM and causes request queuing,
the system either drops KVCache of existing requests,
swaps it out, or migrates it to an available spare GPU to make room for queued requests
(detailed in \textsection{\ref{sec:state-of-the-art}}).
We argue that adjusting KVCache does not fundamentally resolve the queuing
issue caused by memory overloading,
because these methods do not release sufficient memory for all requests, 
i.e., they replace one set of queued requests with another.
Thus, a portion of requests must still be queued, 
still resulting in sharp tail latency increases (\eg{more than 100\,$\times$}).

This paper answers a key question:
\emph{how can we effectively handle the latency spikes caused by
memory overloading in LLM serving?}
To answer this question, we propose a new system mechanism---parameter-centric memory management---to instantly free up
abundant GPU memory upon overloading for all requests to eliminate queuing.
Our method is motivated by two insights.
First,
the HBM usage is dominated by both KVCache and model parameters (34--74\% per GPU, see Table~\ref{tab:model}),
so dropping a portion of parameters can free up sufficient memory
for processing all requests.
While intuitive,
dropping parameters inevitably disrupts the inference process,
making the GPUs with dropped parameters unable to process requests.
Thus, our second insight is that,
due to the massive computational requirements of model serving,
modern LLMs are served with a cluster of GPUs
where the parameters are replicated across multiple GPUs~\cite{bedrock,openai-api,serverless-ray, basten-deepseek,preplaxity-deepseek,llumnix,splitwise}.
As a result, as long as we carefully drop parameters
to ensure complete copies exist cluster-wide,
we can correctly process requests with dropped parameters using cooperative execution.

Our parameter-centric memory management
operates in a three-step process.
First, upon detecting that the serving system
has suffered or is about to suffer from memory overload,
we derive a drop plan and execute it across GPUs to free up sufficient memory.
Afterward, requests executed on GPUs with dropped parameters are seamlessly
rescheduled to groups of GPUs with complete parameters to ensure complete execution.
These requests are executed using parallel inference techniques across GPUs
with pipeline parallelism,
since other techniques like tensor parallelism have more stringent network requirements.
Finally, once the memory demand of the KVCache decreases,
we restore parameters on the original GPUs and reschedule the requests accordingly to achieve
the lowest inference latency.

Although the idea may appear simple, 
achieving parameter-centric memory management necessitates 
tackling a set of challenges.
First,
generating an efficient drop plan should holistically
consider the memory freed up by the dropped parameters
as well as the performance overhead introduced by dropping too many parameters.
Meanwhile, we need a system mechanism to allow existing GPU kernels highly optimized for LLMs
to use the HBM freed up by dropped parameters without modifications.
To this end,
we first leverage the predictable performance pattern of pipeline parallelism---the
more parameters dropped, the more performance overhead incurred---to
quickly derive a drop plan that minimizes the performance overhead
while providing sufficient memory.
Next,
we design a unified GPU virtual memory management system
with advanced GPU virtual memory features~\cite{cuda-memory}
to allow unmodified kernels to access the memory used for parameters
for KVCache (\textsection{\ref{sec:design-memory-manage}}).

Second, efficiently resuming requests after dropping
requires exchanging KVCache between GPUs,
since it is coupled with the parameters.
However,
such an exchange would
significantly interfere with the pipeline-executed requests,
because transferring large KVCache saturates the network used
for forwarding activations.
Observing that the activation transfer is more critical and the network usage is small,
we design a coordinated network transfer engine that prioritizes the activation transfer
to ensure both transfers are not affected (\textsection{\ref{sec:exchange-restore}}).

Finally, 
the pipelined execution across multiple GPUs after parameter dropping
causes GPU bubbles~\cite{sarathi},
resulting in increased serving latencies and degraded throughput.
The throughput degradation is particularly harmful in our setup,
because if requests
are processed at a slower rate, it could lead to another round of memory overloading.
To tackle this problem, we identify the root cause of bubbles
as sub-optimal batch formulation in state-of-the-art systems like Sarathi-Serve~\cite{sarathi}.
By leveraging the observation that under overloading many requests are queued,
we holistically form microbatches of queued requests using a new execution
estimation metric combined with a lookahead batch formulation algorithm.
Our scheduling
minimizes the pipeline bubbles thanks to the holistic formulation during pipelined execution (\textsection{\ref{sec:online-sched}}).

We built {\sys}, the first LLM serving system with
parameter-centric memory management. 
Under various real-world traces and datasets,
when compared with the {\sota} baselines including Llumnix~\cite{llumnix}, vLLM~\cite{vllm} and InferCept~\cite{infercept},
{\sys} achieves up to {12.7-72.2}\,$\times$ tail latency reduction in these workloads,
which further results in {7.2--12.8}\% lower SLO violations
under common SLO factors. 
In summary, this paper makes the following contributions: \\[-10pt]
\begin{itemize}[leftmargin=*,leftmargin=10pt,itemindent=0pt]
    \item A new parameter-centric memory management design for coping with memory overloading under LLM serving (\textsection{\ref{sec:overview}}).  \\[-5pt]
    \item A set of new techniques to make parameter-centric memory management efficient (\textsection{\ref{sec:design}}).  \\[-5pt]
    \item Extensive evaluations confirming the benefits of {\sys} (\textsection{\ref{sec:eval}}).
\end{itemize}

{\sys} is open-sourced at \burl{https://github.com/SJTU-IPADS/kunserve}. 

\section{Background and Motivation}
\label{sec:bg}

\begin{figure}[!t]
        \begin{minipage}{1\linewidth}
        \hspace{-1mm}
        \centering    
        \includegraphics[width=.95\columnwidth, trim=0.25cm 10.7cm 15.98cm 0.25cm, clip]{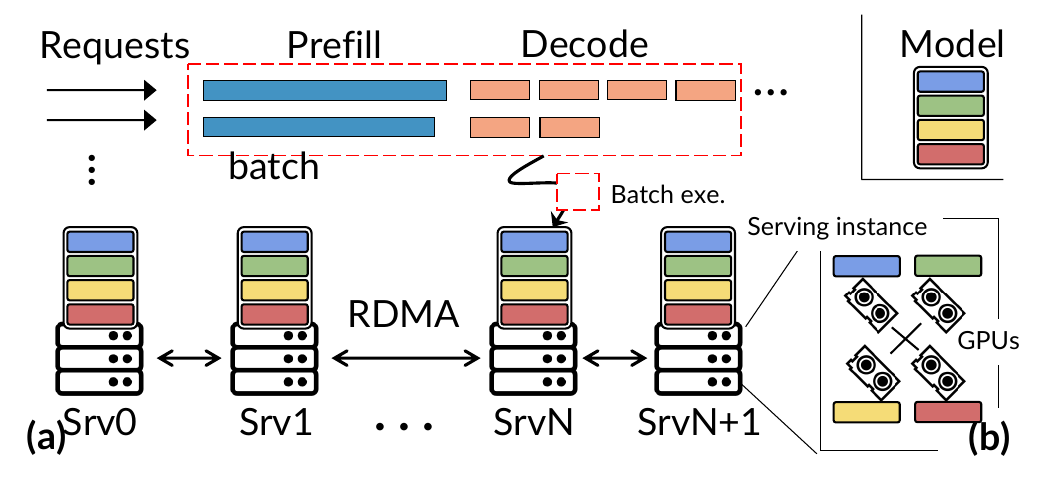} \\[1pt]    
        \end{minipage} \\[3pt]
        \begin{minipage}{1\linewidth}
        \caption{\small{%
            An illustration of a typical LLM serving scenario:
            (a) the model is deployed on different servers with model parallelism 
            and prefill and decode requests are processed in a batched way. 
            \emph{exe.} is abbreviation for execution.
        }}
        \label{fig:bg}
        \end{minipage} \\[-5pt]
        \end{figure}

\subsection{Preliminaries of LLM and LLM serving}
\label{sec:bg-llm}

\nospacestitle{LLM basics. \,}
LLM is a transformer-based~\cite{DBLP:conf/nips/VaswaniSPUJGKP17} deep learning model.
Compared with traditional DNN,
a key difference is that it
executes requests in an \emph{auto-regressive} pattern
with a \emph{prefill} and \emph{decode} phase.
In the prefill phase, the input is fed to the model
to generate the first token of the output.
The decode phase then iteratively generates the rest of the output in a token-by-token way,
where each iteration takes the previously generated token as well as the prefill
input as the context.
The decode\footnote{\footnotesize{We use the term \emph{decode}
to refer to the execution of a single iteration in the decode phase in this paper.}}
ends when the model generates a special end-of-sequence (EOS) token.

During LLM inference, since the same prefix of input is shared across all the iterations,
the internal results (termed \emph{KVCache}) are cached in the GPU memory (HBM)
for acceleration. This makes the computation patterns of prefill and decode different~\cite{splitwise,DBLP:journals/corr/abs-2401-11181,distserve}:
the prefill is compute-bound, while the decode is memory-bound. To improve GPU utilization,
modern LLM inference frameworks fuse prefill and decode requests into a single batch~\cite{sarathi,vllm}.

\stitle{Serving metrics: TTFT and TPOT. \,}
As the output tokens are generated iteratively,
current systems serve requests in a streaming fashion,
i.e., once a token is generated, it is immediately returned to the user.
Thus, both the \emph{prefill latency}
(\textbf{T}ime-\textbf{T}o-\textbf{F}irst-\textbf{T}oken, TTFT)
and the \emph{time to emit each token}
(\textbf{T}ime-\textbf{P}er-\textbf{O}utput-\textbf{T}oken, TPOT)
matter.

\stitle{Deploying LLM instances with parallelism and replication. \,}
LLMs can be deployed on a single GPU or multiple GPUs with parallelism~\cite{DBLP:conf/osdi/LiZZL00HCZGS23,DBLP:journals/corr/abs-1909-08053,DBLP:conf/osdi/ZhengLZZCHWXZXG22}.
Pipeline parallelism (PP) partitions model parameters by layers,
where layers belonging to the same group (\ie{stage}) are executed on the same GPU.
Tensor parallelism (TP) partitions each layer,
while different stages can reside on the same GPU.
Parallelism comes at the cost of extra latency.
For methods with high communication requirements like TP,
parallelism is only applied to GPUs within the same server, because their interconnects are fast.
PP on the other hand, can apply to GPUs across servers thanks to its
ultra-low communication volume.
However, PP suffers from bubbles~\cite{sarathi-v1} especially for requests with a small batch size.
TP and PP can be applied together.

In this paper,
we define the minimal set of GPUs that have a single copy of the model parameters as a \emph{serving instance}.
The GPUs of an instance can be within the same server or across servers,
but typically within the same server for the lowest serving latency unless the model exceeds capacity of a single server, which is rare (\eg{Llama-3-405B}).
For a serving cluster, 
a common practice is to deploy multiple instances with replicated models~\cite{basten-deepseek,preplaxity-deepseek,llumnix,splitwise}, as shown in {\fig{fig:bg}},
because a single instance has limited serving capacity. 

\subsection{TTFT spikes caused by memory overloading}
\label{sec:bg-challenges}

\begin{figure*}[!ht]
    \includegraphics[width=2.1\columnwidth,center]{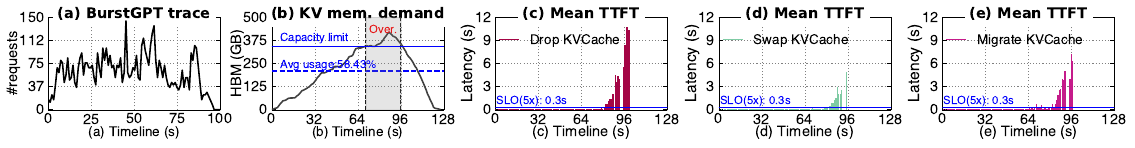} \\[-5pt]
    \begin{minipage}{1\linewidth}
    \caption{\small{{
        Analysis of TTFT increases due to GPU memory overloading (abbreviated as “Over.” in figure).
        (a) The incoming request rate of BurstGPT trace~\cite{burstgpt}. 
        (b) KVCache memory demand on vLLM~\cite{vllm}
        and (c)--(e) requests TTFT of existing solutions (\textsection{\ref{sec:state-of-the-art}}).
    }}}
    \label{fig:motiv}
    \end{minipage} \\[-5pt]
\end{figure*}

\nospacestitle{Huge HBM demands and memory overloading of LLM serving. \,}
The overall memory demand for LLM serving is huge.
For example, when serving a Qwen-2.5-14B model, each token consumes 192\,KB of memory, 
which is already relatively small due to the use of GQA~\cite{DBLP:conf/emnlp/AinslieLJZLS23}, 
a memory-efficient attention mechanism.
A typical burst still introduces an accumulation of 243\,K tokens per GPU
on BurstGPT trace (see \fig{fig:motiv}),
consuming 45\,GB KVCache memory per GPU.

We attribute GPU memory overloading to two causes.
First, real-world traces exhibit spiked loads:
{\fig{fig:motiv}} (a) shows a real-world trace on BurstGPT~\cite{burstgpt},
where the incoming request rate increases by {2\,$\times$} at time {45s} with no clear pattern.
Since the KVCache demand is also proportional to the request rate,
the memory demand can easily exceed the GPU memory capacity.
Second, each request's KVCache may reside in GPU for a long time,
with an unpredictable duration, depending on how long LLMs generate the EOS.
For BurstGPT dataset, the average stay time for a request is 11 seconds, with a variance of 14.9 seconds.
Thus, even the HBM is sufficient to hold incoming requests, 
GPUs still suffer from memory overloading due to the unfinished requests.

{\fig{fig:motiv}} (b) shows how existing serving systems behave under BurstGPT.
During a {640s serving period} (\textsection{\ref{sec:eval-restore}}),
we observed two overloading events on vLLM~\cite{vllm}, a state-of-the-art LLM serving system.
The timing of overloading is strongly related to the request spikes.
Note that we have chosen a practical setup 
where the overall HBM provisioned for KVCache is 2.1\,$\times$ higher than the average requirement.
We use a standard approach~\cite{llumnix} that counts the memory demands 
by considering both the in-processing requests and head-of-line queuing requests.

\stitle{TTFT spikes. \,}
GPU memory overloading severely degrades serving performance.
As shown in {\fig{fig:motiv}} (c),
the TTFT increases significantly after the overloading happens (see (b)).
The increase comes from the queuing delays while waiting for
sufficient memory to be freed.
The queuing time can be lengthy
because the memory can only be freed once the ongoing request batch finishes.
As we have mentioned before, the ongoing requests may take a long time to finish
(\eg{up to {150s} in BurstGPT}).

\subsection{Shortcomings of current solutions}
\label{sec:state-of-the-art}

\begin{figure*}[!t]
        \begin{minipage}{1\textwidth}
        \hspace{-1mm}
        \centering    
        \includegraphics[width=.93\columnwidth, trim=0.25cm 7.5cm 20.2cm 0.25cm, clip]{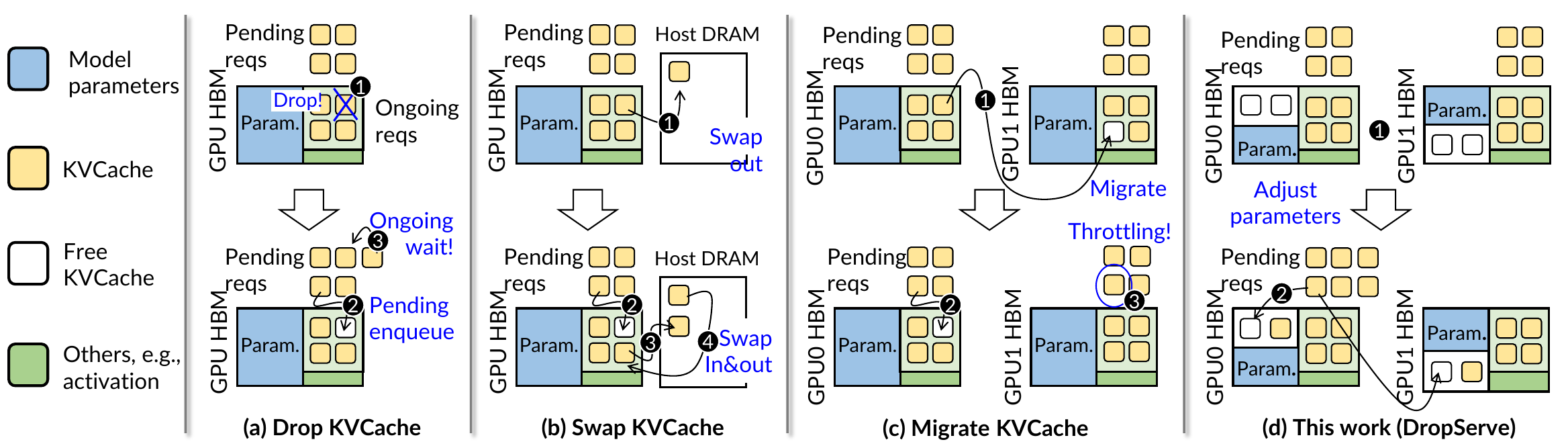} \\[1pt]    
        \end{minipage} \\[0pt]
        \begin{minipage}{1\linewidth}
        \caption{\small{%
        (a)---(c) Existing methodologies to address memory overloading of KVCache.
        (d) How {\sys} tackles this issue 
        via parameter dropping (\ding{182}) and remapping memory to enlarge KVCache region (\ding{183}).         
        }}
        \label{fig:oom}
        \end{minipage} \\[-10pt]
        \end{figure*}

\nospacestitle{Drop the KVCache~\cite{vllm,DBLP:journals/corr/abs-2404-09526,DBLP:journals/corr/abs-2407-00079} (\fig{fig:oom} (a)). \,} 
A straightforward solution is to drop some KVCache of ongoing requests (\ding{182}).
Subsequently, queued requests can be processed with the freed GPU memory (\ding{183}).
However, requests with dropped KVCache must be re-enqueued and recomputed,
which also suffers the queuing overhead (\ding{184})
even without considering the recomputation cost.
As a result, {\fig{fig:motiv}} (c) shows that simply dropping the KVCache
faces up to 239\,$\times$ TTFT increases during memory overloading,
even with a modest average memory load ({56.3\%}).

\stitle{Swap the KVCache~\cite{distserve,vllm,infercept,fastserve} (\fig{fig:oom} (b)). \,}
A classic solution to handle memory overloading is swapping:
when it happens,
the system swaps out the overflowed KVCache
to other storage (e.g., CPU DRAM)
to free the GPU memory for execution (\ding{182}).
The key problem is that as the GPU memory is still insufficient,
there will inevitably be queued requests,
even without considering the swapping overhead.
For example, under overloading, InferCept~\cite{infercept} concurrently swaps
out the KVCache of ongoing requests to hide the transfer overhead,
but the queued requests are still waiting for ongoing requests to finish.
The waiting time can be substantial because the overall decode
time is orders of magnitude higher than TTFT.
As a result,
we still observed a {92}\,$\times$ TTFT spike on InferCept~\cite{infercept}
in {\fig{fig:motiv}} (d).
Worse still, the swapped-out requests (\ding{184}) further suffer high TPOT (see \fig{fig:e2e-latency}).

\begin{table}[t]
    \centering
    \small{
    \resizebox{.99\linewidth}{!}{
    \ra{1.2}        
\begin{tabular}{l|rrr}
        \toprule
        \textbf{Model} &   \textbf{Model size} & \textbf{\#GPU/instance}  & \textbf{Ratio (\%)} \\  \midrule  
        \textbf{Qwen-2.5-14B}     &  {28}\,GB &  1 (80\,GB)            &  34.4              \\
        \textbf{Qwen-2.5-72B}     & {136}\,GB & 4  (320\,GB)           &  42.3              \\
        \textbf{Llama-3.1-405B}   & {756}\,GB & 16 (1,280\,GB)         &  59.1              \\
        \textbf{Qwen-3-235B}        & {479}\,GB & 8 (640\,GB)            &  74.8              \\
        \textbf{DeepSeek-V3-671B}      & {1,572}\,GB & 32 (2,560\,GB)            &  61.4             \\
        \bottomrule
        \end{tabular}
    }
    } 
    \begin{minipage}{1\linewidth}
    \caption{\small{{Popular LLM models, their parameter memory usage, the number of GPUs
        belonging to an instance, and the parameter memory usage ratio.
        Note that within an instance, Qwen-3-235B and DeepSeek-V3-671B are configured with expert parallelism with degrees 8 and 32,
        respectively, a common serving setup~\cite{DBLP:journals/corr/abs-2412-19437}.
    }}}
    \label{tab:model}
    \end{minipage} \\[-5pt]
\end{table}

\stitle{Migrate the KVCache~\cite{llumnix} (\fig{fig:oom} (c)). \,} 
Finally, observing that a serving cluster typically has multiple instances,
a recent work (Llumnix~\cite{llumnix}) migrates requests from a memory-overloaded GPU to other (relatively)
spare GPUs (\ding{182}) for pending requests (\ding{183}).
The observation is that while no single GPU can hold all the pending requests,
we can migrate requests to reduce fragmentation to free up sufficient memory.
However, the queued requests can still be stalled because memory is occupied by
migrating requests or the destination node is also memory-overloaded (\ding{184}).
Worse still, under spike workloads,
there is little room for using migration to free up memory
because the overall memory KVCache is insufficient even without considering fragmentation.
Thus, as shown in \fig{fig:motiv} (e), migration still leads to a
148\,$\times$ P99 TTFT increase (compared to the P50).

\section{System Overview}
\label{sec:overview}

\nospacestitle{Approach: online parameter dropping. \,}
As mentioned in the introduction,
{\sys} is based on two key observations of LLM serving:
(1) parameters typically take up a considerable portion of HBM per GPU
(see Table~\ref{tab:model}) that can be used for KVCache
and
(2) parameters are replicated across instances so dropping them for KVCache
does not impact LLM serving.
\fig{fig:oom} (d) illustrates {\sys}'s main approach and a
comparison with other baselines assuming two instances and each instance uses one GPU. 
When the HBM used for KVCache is exhausted on GPU0 and GPU1,
we instantly drop the second half of layers on GPU0 and 
the first half of layers on GPU1 (\ding{182}).
Then, the queued requests are rescheduled on both GPUs (\ding{183})
for execution via pipeline parallelism. 

\stitle{Discussion: why pipeline parallelism? \,}
We chose pipeline parallelism because
the network requirement can be easily satisfied
with the interconnects between instances.
Specifically,
it requires orders of magnitude smaller
communications than other parallelism setups that support
execution after the parameter drop like tensor parallelism.
While instances could link together via fast interconnects like NVLink for tensor parallelism,
the domain of NVLink is much smaller than networks that could serve
pipeline parallelism well like RDMA~\cite{nvidia-compute-network}.
Thus, under overloading, we may be unable to find sufficient
instances connected by NVLink.

\begin{figure}[!t]
        \begin{minipage}{1\linewidth}
        \hspace{-1mm}
        \centering    
        \includegraphics[width=.8\columnwidth, trim=0.25cm 11.6cm 20.5cm 0.25cm, clip]{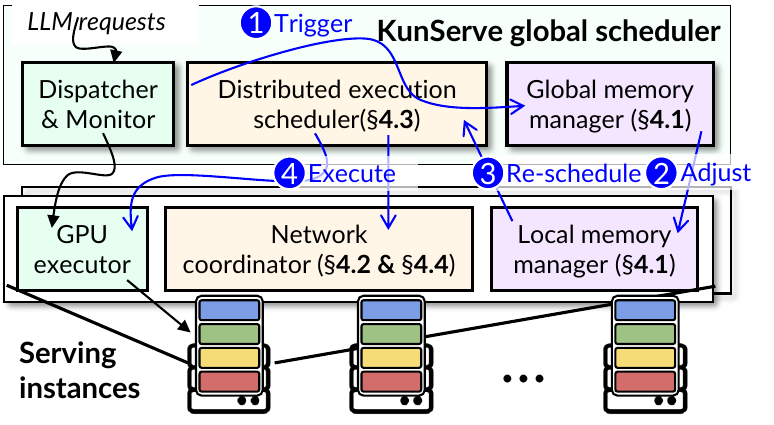}
        \end{minipage} \\[1pt]
        \begin{minipage}{1\linewidth}
        \caption{\small{%
            System overview of {\sys}.
        }}
        \label{fig:sys-arch}
        \end{minipage} \\[-10pt]
        \end{figure}

\stitle{System architecture. \,}
\fig{fig:sys-arch} illustrates our system architecture 
as well as the workflow of parameter-centric memory management
for handling memory overloading.
{\sys} is a cluster-serving system that manages a set of LLM serving instances. 
Requests are routed through a global dispatcher, which enqueues them to the local executor 
of each instance for execution. Our dispatcher incorporates the load-balancing design 
from Llumnix~\cite{llumnix}. The global monitor collects usage information and calculates 
the load metric for each instance.

Once a memory overloading event is detected by the monitor,
it invokes our global memory manager (\ding{192}) to generate dropping plans.
The plan is then forwarded to the local manager 
on the involved instances (\ding{193}) 
to adjust the memory according to the plan
(details in \textsection{\ref{sec:design-memory-manage}}). 

{After parameter dropping, {\sys} re-scheduled 
queued requests and ongoing requests to execute
on instances with enlarged memory using pipelined parallelism (\ding{194}).}
To ensure a smooth resumption of the
requests whose KVCache is not on the target instances
to avoid computation waste,
our network coordinator
exchanges the KVCache of ongoing requests
between instances without blocking the activation transfer of 
pipelined execution (\textsection{\ref{sec:exchange-restore}}).
Meanwhile,
our optimized pipelined scheduling minimizes the bubbles in the
upcoming execution (\textsection{\ref{sec:online-sched}}).

Finally, once the memory demand goes down,
{\sys} dynamically restores parameters such that future requests can
execute with lower latency. (\textsection{\ref{sec:design-others}}).

\section{Detailed Design and Implementation}
\label{sec:design}

\begin{figure}[!t]
        \begin{minipage}{1\linewidth}
        \hspace{-1mm}
        \centering    
        \includegraphics[width=\columnwidth]{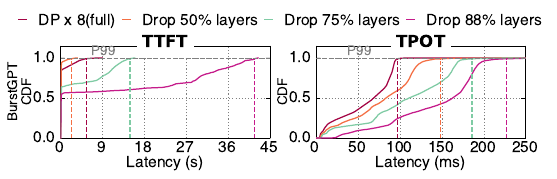} \\[1pt]    
        \end{minipage} \\[-5pt]
        \begin{minipage}{1\linewidth}
        \caption{\small{%
            A comparison of the latency of different parallelism on BurstGPT
            dataset. All setups are evaluated with 8 GPUs.
        }}
        \label{fig:pp-analysis}
        \end{minipage} \\[-5pt]
\end{figure}

\subsection{Parameter drop during memory overloading}
\label{sec:design-memory-manage}

\noindent
Upon overloading, 
{\sys} needs to generate a drop plan to free up sufficient memory. 
Besides the memory requirement, 
the plan has to meet the following requirements: 
(1) we need to generate the plan quickly online, 
(2) the plan needs to ensure a correct execution and 
(3) the plan needs to minimize the performance loss 
caused by parameters drop. 

For (2),
we only need to ensure that all the instances combined have a complete copy of parameters.
However,
dropping too many parameters incurs a performance cost.
For example, suppose we are serving a 7-layer model with 7 instances.
While dropping 6 layers on all instances can free 85\,\% of the HBM for KVCache,
it forces the scheduler to split the batch into microbatches with smaller sizes,
reducing the GPU batch execution efficiency~\cite{flashinfer} and
making the system more vulnerable to pipeline bubbles.
{\fig{fig:pp-analysis}} compares the serving latencies for
different degrees of parameter dropping.
We can clearly see that the more parameters are dropped, the higher the execution latency.

\begin{figure}[!t]
        \begin{minipage}{1\linewidth}
        \hspace{-1mm}
        \centering    
        \includegraphics[width=1\columnwidth, trim=0.25cm 5.35cm 17.8cm 0.25cm, clip]{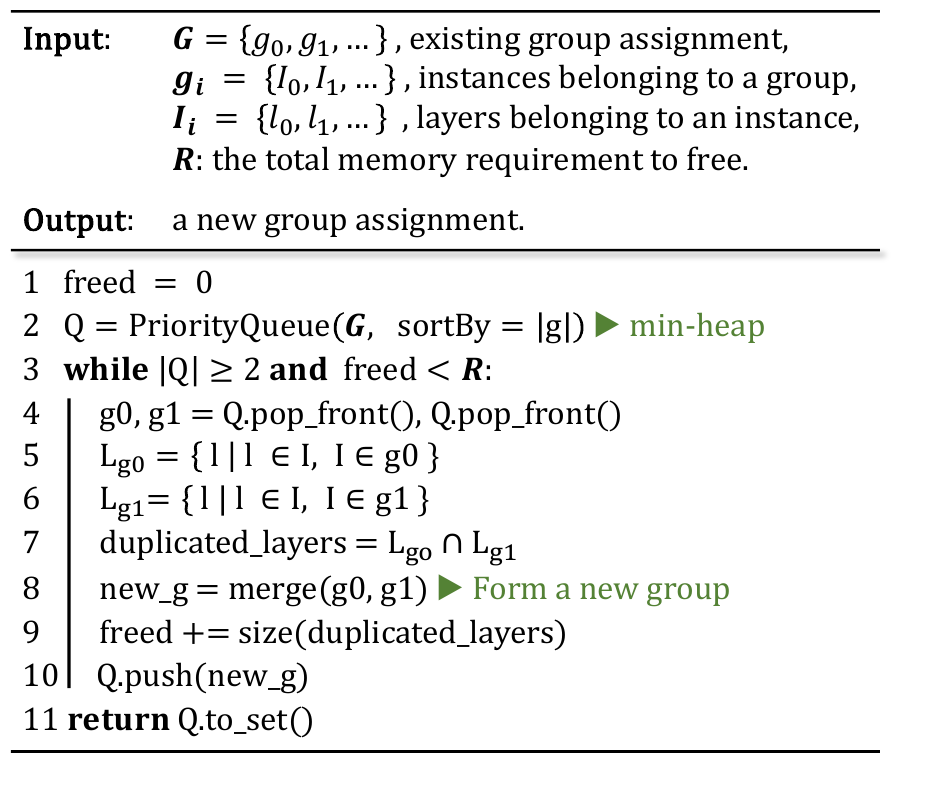}
        \end{minipage} \\[-10pt]
        \begin{minipage}{1\linewidth}
        \caption{\small{%
            The pseudocode of drop plan generation algorithm. 
        }}
        \label{alg:param-drop}
        \end{minipage} \\[-10pt]
        \end{figure}

A key takeaway from {\fig{fig:pp-analysis}} is that
the performance loss is strongly correlated with the number of instances involved in processing
a request, i.e., pipeline stages.
Thus, we design a greedy-based parameter dropping algorithm
by grouping as few instances as possible to minimize performance loss. 

Algorithm~\ref{alg:param-drop} shows the details of our method
that groups instances into groups to free up memory.
The initial configurations ($G$) follow the setups without a drop, 
e.g., each instance itself is a group. 
To support greedy grouping, the group records the number of instances 
involved ($g_i$) and all instances are stored in a priority queue ($\mathcal{Q}$). 

Upon overloading, we first compute the memory demand of all queued requests ($R$)
and enter line 1.
Afterward, we iteratively group instances
and then drop parameters to free more space (lines 3--9).
For example, if there are three groups with sizes of 1, 2, and 3,
we will select the two groups with sizes of 1 and 2 to form a new group (lines 5--6).
For the selected groups,
we drop a copy of the redundant parameters (line 7)
and update the available memory (line 9).
At the end of the iteration, the selected two groups are merged into a new group
and inserted back into the priority queue (line 8).

The iteration continues until the memory requirement is satisfied or
it fails to find a drop plan (line 3).
In case we cannot find a plan, we fallback to the KVCache-centric solution
to ensure continuous execution and autoscale the instance numbers.
The complexity of the plan generation is {$O(N\log{N})$},
so we can quickly execute it online even with a large number of instances.

\begin{figure}[!t]
        \begin{minipage}{1\linewidth}
        \hspace{-1mm}
        \centering    
        \includegraphics[width=1\columnwidth, trim=0.25cm 13cm 14.1cm 0.25cm, clip]{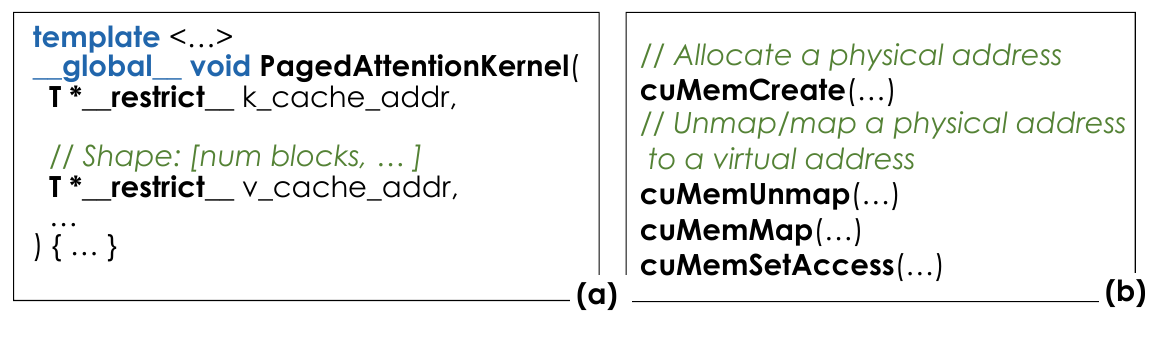}
        \end{minipage} \\[-5pt]
        \begin{minipage}{1\linewidth}
        \caption{\small{%
            (a) The GPU kernel signature of the pagedattention kernel~\cite{vllm-code}. 
            (b) CUDA virtual memory management APIs~\cite{cuda-memory}.
        }}
        \label{fig:pagedattention}
        \end{minipage} \\[-10pt]
        \end{figure}

\stitle{Local instance memory management. \,}
A key challenge of executing the drop plan at each instance
is how to allow existing attention kernels to use the freed parameter memory.
As shown in {\fig{fig:pagedattention}} (a),
the kernels are written with a single static memory layout,
e.g., \texttt{[kcache_addr, kcache_addr + num_blocks * block_size]},
not multiple virtual memory ranges provisioned dynamically.
One possible solution is to rewrite these kernels to suit the
new memory layout.
However, efficiently rewriting LLM kernels is non-trivial
due to the complex and evolving nature of LLM kernels.
Simple rewrites lead to performance drops that
require months of iterative development to optimize~\cite{DBLP:journals/corr/abs-2405-04437}.

To tackle the problem, 
we observe that recent GPUs have introduced application-controlled virtual memory management APIs:
as shown in {\fig{fig:pagedattention}} (b).
For example, \texttt{cuMemCreate} allows allocating a piece of GPU physical memory
and \texttt{cuMemMap} can map it to an arbitrary virtual address.
With such APIs, we can dynamically change the virtual address space of KVCache 
without modifying the kernel code.
The overhead of calling these APIs is in the microsecond level (5\,ms on our platform),
which is negligible to the LLM inference time. 
Specifically, our local instance memory management
holistically manages the GPU physical memory
for both the parameters and the KVCache with \texttt{cuMemCreate}.
Afterward, when executing the drop plan received from the global coordinator,
we first identify the physical memory of the dropped parameters.
Then we extend the memory for KVCache
by mapping the tail of the KVCache memory to the freed physical memory
with \texttt{cuMemCreate}.

\subsection{Smooth transition of requests from undropped to
dropped states with coordinated KVCache exchange}
\label{sec:exchange-restore}

\noindent
Because the KVCache has a one-to-one mapping with the model parameters,
we cannot simply execute ongoing decode requests due to the lack of KVCache.
For example, suppose a request has executed on instance A,
and A has formed a group with instance B due to memory overloading.
After the drop, A will only have parameters of layers 0--4,
while B will have layers 5--7.
Hence, B cannot directly execute the 5--7 layers of a request originally on A
because the required KVCache is on A.
Similarly, A cannot execute the 0--4 layers of a request originally on B.
One intuitive solution is to recompute the KVCache on B.
This is expensive since it causes queued requests to wait for the recomputation
even without considering the recomputation time.

\stitle{Network-based KVCache exchange. \,}
We choose to exchange the KVCache through the network
to avoid recomputation.
The KVCache is exchanged because after A and B have formed a group,
ongoing requests on A need to transfer their KVCache to B,
while B needs to do the same vice versa.
A drawback of the exchange is that the requests with the exchanged KVCache will be
stalled during the exchange,
which we found to be acceptable in practice.
This is because the network
between instances such as RDMA is sufficient for transferring the KVCache quickly.
For example, KVCache exchange typically introduces {1--2\,s} stall time on our {200\,Gbps network}.
This means a 10\,ms increase at most in the TPOT metric of a response with 200 decode tokens.

Note that during the stall, we can still schedule new requests
queued due to memory overloading to fully utilize the GPUs. 
While in principle we can leverage techniques like attention offloading 
(also called model-attention disaggregation)~\cite{DBLP:journals/corr/abs-2405-01814}
to concurrently execute stalled requests during the KVCache exchange,
we found the excessive complexity of the implementation is not worth the effort.

\stitle{Coordinated KVCache exchange. }
Although straightforward, KVCache exchange could block new request if not implemented properly,
because the exchange competes for bandwidth with activation transfers in pipelined execution.
Since the exchange time is much longer than forwarding the activation,
When the activation is waiting for the exchange to finish,
it will leave the GPUs idle, causing non-negligible performance loss.
Observing that the activation transfer is much smaller yet more critical,
we design a coordinated exchange mechanism to prioritize the activation transfer.
Specifically, we transfer KVCache in finer-grained chunks 
such that the transferring a chunk takes similar time to executing a pipeline stage. 
After transferring one chunk, we will check whether there 
will be activation transfer.
If so, we pause the KVCache transfer and let the activation transfer go first.

\subsection{Serving requests efficiently after parameter drop}
\label{sec:online-sched}

\begin{figure}[!t]
        \hspace{-2.5mm}
        \begin{minipage}{1\linewidth}
        \centering    
        \includegraphics[width=1.05\columnwidth, trim=0.25cm 11.6cm 15.6cm 0.25cm, clip]{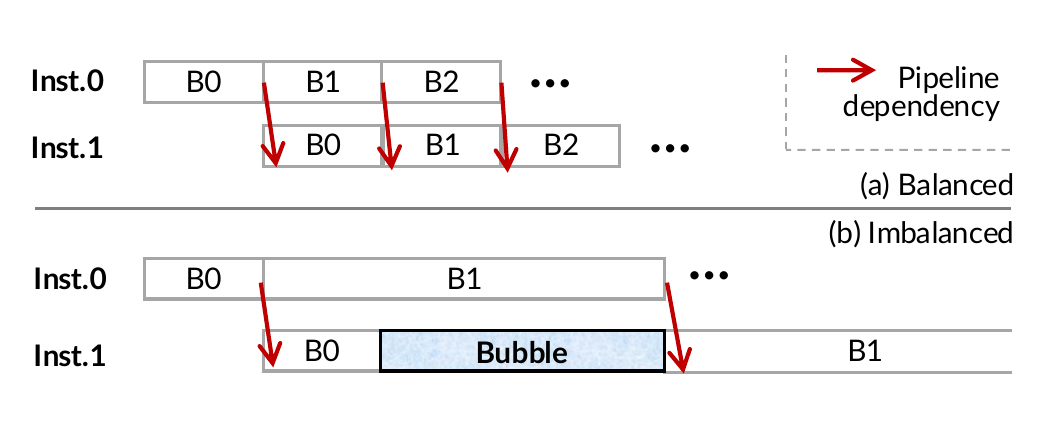} 
        \end{minipage} \\[-8pt]
        \begin{minipage}{1\linewidth}
        \caption{\small{%
        An illustration of pipeline execution bubbles caused by
        imbalanced execution time of microbatches.
        }}
        \label{fig:pp-bubble}
        \end{minipage} \\[-5pt]
        \end{figure}

\begin{figure}[!t]
    \hspace{-3mm}
    \begin{minipage}{1\linewidth}
    \centering    
    \includegraphics[width=1.02\columnwidth, trim=0.25cm 4.2cm 16.43cm 0.25cm, clip]{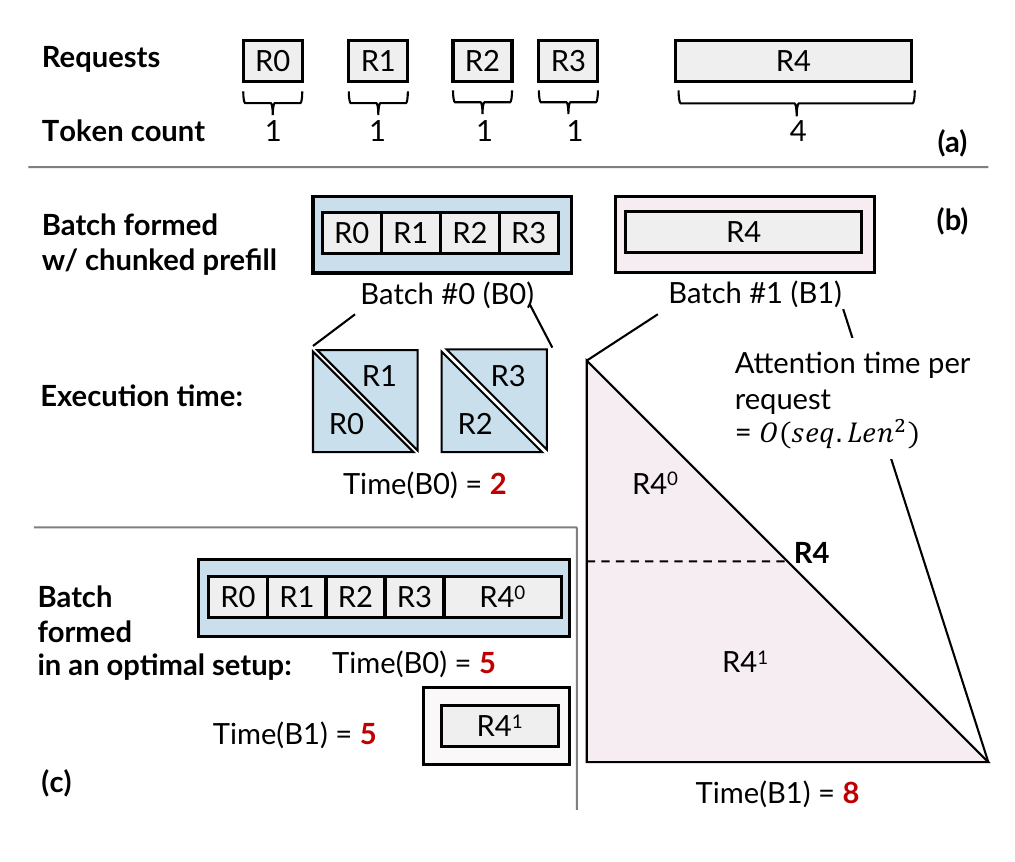} 
    \end{minipage} \\[-8pt]
    \begin{minipage}{1\linewidth}
    \caption{\small{%
    (a) An illustration of serving requests to execute.
    (b) The imbalanced batch execution time of existing chunking method.
    (c) A balanced formulated batch configuration.  
    }}
    \label{fig:pp-imbalance}
    \end{minipage} \\[-15pt]
    \end{figure}

\nospacestitle{Key problem: pipeline bubbles caused by unbalanced microbatch execution time. \,}
A problem of pipeline execution after parameter drop
is that the system suffers from degraded throughput due to pipeline bubbles.
The bubbles arise from the imbalanced execution time of different
microbatches, as illustrated in {\fig{fig:pp-bubble}} (b).
For example, when B1's execution time is longer than B0,
Inst.1 must wait for B1 to finish before it can execute the layers on B2.

\stitle{A preliminary on the state-of-the-art pipeline microbatch formulation.  \,}
Modern pipeline implementations rely on chunked prefill to reduce pipeline bubbles.
Specifically, they~\cite{sarathi,vllm} form microbatches
in a token-count-based manner,
which balances the execution time of different microbatches
by ensuring each microbatch has a similar number of tokens.
As shown in {\fig{fig:pp-imbalance}} (a),
suppose 5 requests (R0--R4) arrive at an instance in turn,
and the budget for each microbatch is 4 tokens.
The scheduler first merges incoming requests into one microbatch (R0--R3 in (b)).
R4 itself forms another microbatch (B1).
Note that if R4 exceeds the budget, the scheduler will chunk it
into two segments for execution.

\stitle{Inefficiency of token-count-based chunking. \,}
A key issue is that the microbatch execution time
is not linearly proportional to the total token count,
because the attention computation of each request is quadratic to its token count,
as shown in {\fig{fig:pp-imbalance}} (b).
Moreover,
if a request is chunked into two parts,
the latter chunk is slower than the former even when the tokens are the same,
because the latter chunk has to additionally compute the attention with the former chunk.

\begin{figure}[!t]
        \hspace{-2.5mm}
        \begin{minipage}{1\linewidth}
        \centering    
        \includegraphics[width=0.9\columnwidth, trim=0.25cm 12.8cm 23.3cm 0.25cm, clip]{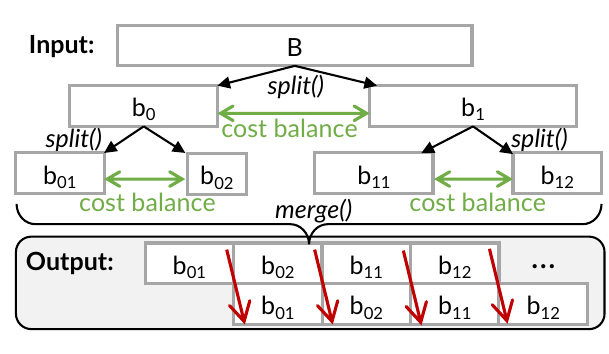} 
        \end{minipage} \\[-8pt]
        \begin{minipage}{1\linewidth}
        \caption{\small{%
        An illustration of how lookahead batch formulation
        recursively generate balanced microbatches. 
        }}
        \label{fig:lookahead}
        \end{minipage} \\[-15pt]
        \end{figure}

\stitle{The lookahead batch formulation. \,}
Fortunately,
under bursts, we have sufficient requests queued.
Thus, 
we can re-form the microbatches across them 
by looking ahead at all requests queued.
To efficiently find the balanced microbatch configuration,
we propose a heuristic divide-and-conquer algorithm. 

\begin{figure}[!t]
        \vspace{2mm}
        \begin{minipage}{1\linewidth}
        \hspace{-1mm}
        \centering    
        \includegraphics[width=1\columnwidth, trim=0.25cm 4.2cm 17.8cm 0.25cm, clip]{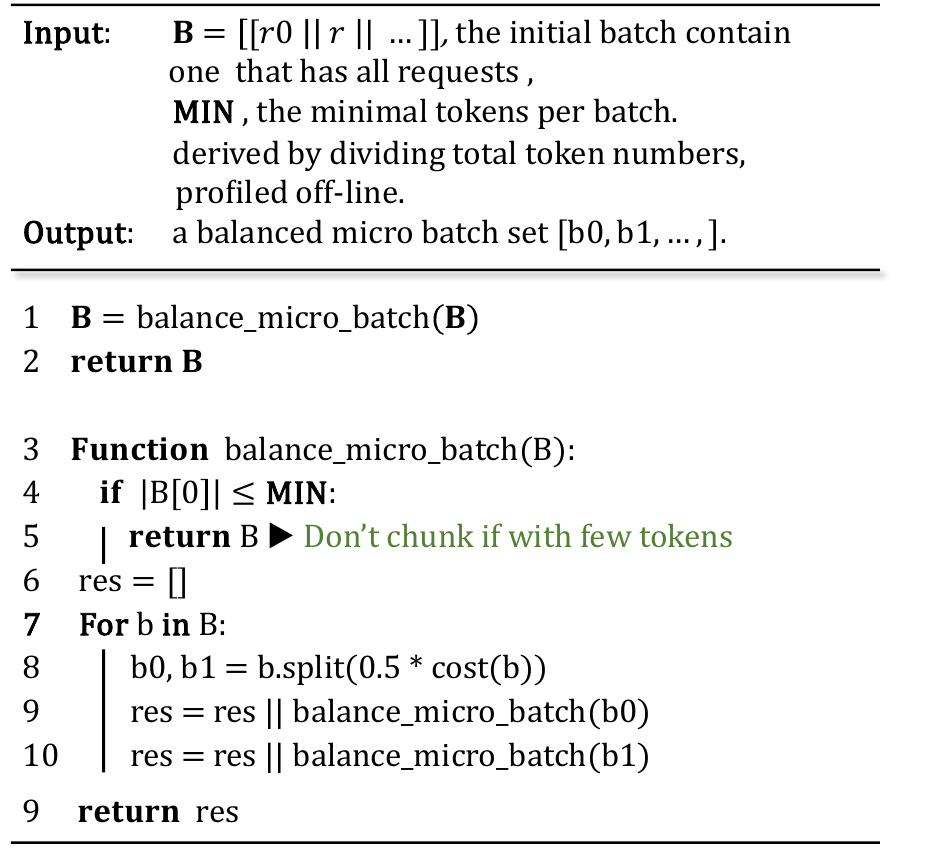}
        \end{minipage} \\[-5pt]
        \begin{minipage}{1\linewidth}
        \caption{\small{%
            The pseudocode of the divide-and-conquer microbatch formulation algorithm.
        }}
        \label{alg:lookahead}
        \end{minipage} \\[-10pt]
        \end{figure}

Our method works in two steps.
First, we adopted a retrofitted cost model 
to precisely estimate the execution time of a microbatch.
Second, we recursively generate the microbatch configurations
according to the cost model.
Specifically, balancing can be done
by looking ahead all tokens to be chunked in a recursive manner,
as shown in {\fig{fig:lookahead}}. 
The initial batch contains a single microbatch with all tokens,
which is then recursively split into two cost-balanced 
microbatches until it reaches a balanced setup. 

\fig{alg:lookahead} shows the detailed pseudocode.
The algorithm complexity is $O(\log{L})$
so it can be quickly solved online. 
For simplicity, we omit the details of split,
which divides requests in a batch into chunks and
returns a new microbatch set whose aggregated cost is equal to the objective
($0.5 \times cost(b)$).
This ensures that each microbatch has sufficient tokens
to fully utilize the GPU. 
One thing to note is that the generation halts
once the number of tokens to form a batch is below a threshold (line 4--5).

A key to the effectiveness of the above algorithm is to
accurately estimate the execution time (\ie{cost}) of a microbatch.
We derive the cost model using a bottom-up approach:
we first model the cost of executing a chunk of a request,
 then we sum the cost of all chunks in a microbatch as its cost. 
Specifically, 
suppose we have a microbatch set $\mathcal{B}$,
denoted by \( \mathcal{B} = \{b_1, b_2, \ldots, b_m\} \),
The chunks are chunked from 
a request set of size \( n \),
denoted by \( \mathcal{R} = \{r_1, r_2, \ldots, r_n\} \).
The cost of a chunk \( c_{ij} \), {\( \text{cost}_{c_{ij}} \)},
can be formulated as follows:

\begin{equation}
    \text{cost}_{c_{ij}} = \alpha\, 
    \left( \overbrace{p_{ij}c_{ij}}^{\text{prefix-attn}} + 
    \overbrace{\frac{c_{ij}^2 + c_{ij}}{2}}^{\text{self-attn}} \right) + 
    \beta\, \overbrace{c_{ij}}^{\text{FFN}} + 
    \overbrace{\gamma}^{\text{other}}
    \label{eq:req}
\end{equation}

\noindent 
The equation consists of four parts: the cost to compute attention
with previous tokens (\textbf{prefix-attn}); the cost to compute attention
with the chunk itself (\textbf{self-attn});
the cost of computing the activations (\textbf{FFN} (Feed-Forward Network)) for tokens;
and others. 
The prefix tokens of each chunk can be calculated as
$p_{ij} = \sum_{k=1}^{j-1} c_{ik}$.
The \textbf{prefix-attn} and \textbf{self-attn} models
the quadratic cost of attention computation missed by existing models,
e.g., NanoFlow~\cite{nanoflow} does not consider \textbf{self-attn}, while DistServe~\cite{distserve} does
not take \textbf{prefix-attn} into account.

Our model depends on several hyperparameters (e.g., $\alpha$) that can be determined through offline profiling:
before the system is deployed for serving, we run multiple inference samples offline,
collect their execution times, and then use
the least squares method~\cite{least-square} to determine all hyperparameters.

Given the cost of each chunk,
we can sum all the costs of chunks in a microbatch to get the cost of the microbatch:
\begin{align}
    & b_k = \left\{ c_{ij} \mid x_{ij} = k  \land c_{ij} > 0 \right\}, \quad \forall k \in \{1, \ldots, m\} \\
    & \text{cost}_{b_k} = \sum_{\substack{(i,j) \\ x_{ij} = k}} \text{cost}_{c_{ij}} {- (|b_k| - 1) \gamma}
\end{align}

\noindent 
Note that the term $-(|b_k|-1)\lambda$ reflects the elimination of duplicated
parameter-loading when executing a batch,
as requests in a batch share the same model parameter.
Like other hyperparameters, $\lambda$ can be fitted with offline profiling.

Empirically, our cost model accurately models the execution time of a microbatch
for common sequence lengths in \fig{fig:cost-model}.
As a result, the pipelined execution with our
lookahead formulation can significantly reduce the execution bubbles
(see \fig{fig:ablation-net}).

\stitle{Discussion: the generality of lookahead batch formulation and cost model. \,}
While in principle, we could also apply lookahead batch formulation
to general LLM serving with pipeline execution,
it has one obstacle that the formulation assumes a sufficient number of requests
queued to ``lookahead'' to be effective.
Under normal serving without bursts,
waiting for requests to be looked ahead may add additional latency,
which we leave possible solutions as a future work.

Besides, readers may findEq.~\ref{eq:req}
still has a part that has a linear correlation with the number of tokens (FFN),
so if the cost is dominated by FFN, existing token-count-based cost models may suffice.
We argue that our retrofitted cost model is still important because
the quadratic terms (prefix-attn and self-attn) would become significant when the token count increases (e.g.,
for requests with more than 4K tokens, which are common in real-world workloads~\cite{longbench}, see \textsection{\ref{sec:eval-setup}}),
so existing works can leverage our model for a more accurate estimation of microbatch execution time.

\subsection{Dynamic restore and fault tolerance}
\label{sec:design-others}

\nospacestitle{Dynamic parameter restoration.}
While dynamic parameter drop described in \textsection{\ref{sec:design-memory-manage}}
can free up memory for new requests under memory overloading,
the pipelined execution is not optimal under normal execution
because (1) pipelined execution suffers from more frequent weight loading 
and (2) it has bubbles.
Normal execution cannot simply apply our {lookahead} scheduling
described in \textsection{\ref{sec:online-sched}}
because there are insufficient numbers of requests to balance.

To this end, {\sys} dynamically restores parameters to return to a normal
non-pipelined execution once the overloading fades away.
Specifically, when the monitor detects that the total
KVCache usage is below a threshold,
it triggers a restoration process by loading the dropped parameters
back to the GPUs.
Currently we use a simple threshold where the memory usage
is below {50}\,\% of the GPU (without drop).
The missing parameters are pulled from instances whenever possible
using the network between instances.

Two things need to be noted about the restoration.
First,
we overlap restoring with the normal request processing.
Second, since {\sys} is concurrently restoring when the request is executing,
the parameter pulling process may block activation transfer
of normal requests, causing latency increases (see \fig{fig:ablation-net}).
Thus,
we adopted a similar coordinated network transfer
approach described in \textsection{\ref{sec:exchange-restore}}
to ensure a smooth execution of pipelined requests by
prioritizing the pipeline network over the parameter transfer.

\stitle{Fault tolerance. }
Unlike traditional LLM serving where failures between instances are isolated,
a failure node in {\sys} can disrupt other instances that are involved in the same
pipeline-parallel group.
Thus, we dynamically restore these affected instances to
ensure normal execution under failures.
By replicating parameters in host DRAM or SSDs, we can always ensure successful parameter 
restoration.

\section{Evaluation}
\label{sec:eval}

\subsection{Experiment setup}
\label{sec:eval-setup}

\nospacestitle{Testbed. \,} 
We evaluate {\sys} on two clusters listed in Table~\ref{tab:cluster}.
Cluster A has one GPU per server so it is typically used for 
running small models (e.g., 14\,B models). 
Cluster B has multiple GPUs per server interconnected 
with fast NVLink, so it is suitable for running larger models (e.g., 72\,B models)
with tensor parallelism.

\stitle{Evaluated models. \,}
Similar to prior works~\cite{sarathi,splitwise,distserve},
we choose open-source models with leading accuracy:
Qwen-2.5-14B and Qwen-2.5-72B~\cite{qwen2.5}.
Both models adopt GQA~\cite{gqa} to reduce KVCache size while maintaining high accuracy.
We do not choose models with huge KVCache usage (e.g., models with MHA~\cite{DBLP:conf/nips/VaswaniSPUJGKP17})
that could easily exhaust GPU memory---though {\sys} is more effective when serving such models.
This is because these models are being replaced by more KVCache-efficient variants.
Table~\ref{tab:model} lists instance configurations of each model.
For the 72B model, we use tensor parallelism to serve requests on multiple GPUs.


\stitle{Evaluated traces and datasets. \,} 
Since memory overloading is sensitive to the request arrival pattern,
we use a real-world trace BurstGPT~\cite{burstgpt} with known 
request arrival information (i.e., the invocation time of each request)
as our main evaluated application. 
Following the guide of BurstGPT, we scale BurstGPT's RPS to fit the serving capacity
of our testbed using a scaling method that preserves the temporal pattern of the trace.
Specifically, we upscale the trace with TraceUpscaler~\cite{traceupscaler},
and ensure that the average memory demand is lower than 60\% of the total memory
during the entire evaluation of the trace.

Besides the arrival pattern,
LLM serving is also sensitive to the input and output length of requests.
Thus, given the trace, we further evaluate requests from representative
datasets representing different scenarios,
similar to prior works~\cite{spotserve,DBLP:conf/osdi/LiZZL00HCZGS23}:  \\[-15pt]
\begin{itemize}[leftmargin=*]
    \itemsep0.5em    
    \setlength{\itemindent}{0em}     
    \item \textbf{BurstGPT.} It is the original dataset of BurstGPT~\cite{burstgpt},
    representing a conversion workload so both TTFT and TPOT are important.
    The average input and output lengths are {642} and {262}, respectively. \\[-15pt]

    \item \textbf{ShareGPT.} 
    ShareGPT~\cite{sharegpt} is another popular chatbot dataset
    that is widely evaluated on~\cite{llumnix,distserve,loongserve,sarathi}.
    Its input and output lengths are longer than BurstGPT,
    representing a workload that is more sensitive to GPU memory provisioning.
    The maximal input length is {4K}, and
    the average input and output lengths are 1,660 and 373, respectively.
    Like BurstGPT, low TTFT and TPOT are both important for benchmark using this dataset. 
    \\[-15pt]

    \item \textbf{LongBench.} 
    LongBench~\cite{longbench} is another popular dataset used for 
    evaluating document summarization tasks~\cite{distserve},
    e.g., summarizing news, articles and
    scientific papers.    
    The average input length is {5.9\,K} and the average output length is {499}.
    Since the user expects a quick response to the summarized content, 
    TTFT is also important. 
\end{itemize}
\noindent

\begin{figure*}[!t]
        \begin{minipage}{1\linewidth}
        \hspace{-1mm}
        \centering    
        \includegraphics[width=\columnwidth]{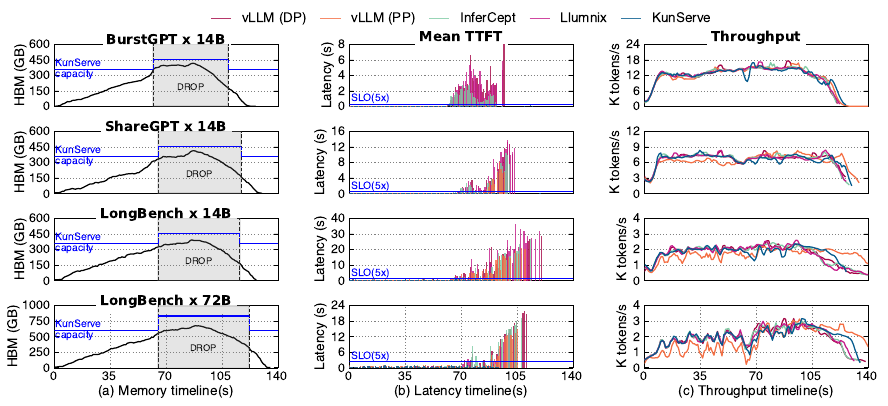} \\[1pt]    
        \end{minipage} \\[-5pt]
        \begin{minipage}{1\linewidth}
        \caption{\small{%
        {
            First column: the memory usage pattern of {\sys}.
            Second column: the mean TTFT during the evaluation.
            Third column: the throughput during the evaluation.
        }}}
        \label{fig:e2e}
        \end{minipage} \\[-10pt]
\end{figure*}

\begin{table}[!t]
    \centering
    \small{
        \resizebox{.97\linewidth}{!}{
            \ra{1.2}

            \begin{tabular}{lrr} \toprule
                                 & \textbf{Cluster A} ($s$ x $g$)     & \textbf{Cluster B} ($s$ x $g$)     \\ \hline
                GPU              & A800 80\,GB (8x1)           & H800 80\,GB (2x8)         \\
                GPU-GPU (scaleup)          & N/A              & 300\,GB/s NVLink    \\
                GPU-GPU (scaleout)          & 200\,Gbps RDMA       & 400\,Gbps RDMA    \\ \bottomrule
                \end{tabular}        
            
        }
    } \\[0pt]
    \begin{minipage}{1\linewidth}
        \caption{\small{{
        Testbed. $s$ is the number of servers and 
        $g$ is the number of GPUs per host. 
        The scaleup and scaleout here means scale-up network
        and scale-out network, respectively.
        The reported bandwidth is unidirectional.
        }}}
    \label{tab:cluster}
    \end{minipage} \\[-5pt]
\end{table}

\stitle{Baselines. \,}
We compared with the state-of-the-art LLM serving systems
with various techniques to cope with memory overloading.
For all systems,
we have carefully tuned their configurations to meet the
optimal performance without memory overloading.
We have also enabled all known serving optimizations to these
systems even though the vanilla systems are not optimized (e.g., InferCept~\cite{infercept}).
For those with our optimizations, we have calibrated that
our optimizations enabled better performance than the original open-sourced codebase.
More specifically, our baselines are:  \\[-12pt]
\begin{itemize}[leftmargin=*]
    \itemsep0.5em    
    \setlength{\itemindent}{0em}     
    \item \textbf{{{vLLM}} (default + PP)~\cite{vllm}.} 
    We compare two configurations of vLLM (release v0.6.3):
    The default configuration stores the entire parameters on each instance,
    while pipelined parallelism (PP) further frees half of the parameters on each instance
    and leverages PP to execute requests across two instances.
    This setup frees up more memory for KVCache, but it also introduces
    pipelined execution overhead.
    By default, vLLM uses recomputation to cope with memory overloading. 
    We compared the vLLM with swapping to InferCept described below.

    Before the evaluation, we carefully tuned the configurations of vLLM.
    Specifically, we tuned the block size to achieve the best performance
    under our setup.
    We chose 64 because (1) it is small enough to avoid memory fragmentation
    while (2) it is sufficiently large to achieve good performance~\cite{flashinfer}.
    \\[-10pt]

    \item \textbf{InferCept~\cite{infercept}.} 
    InferCept designs an optimized swap mechanism that eliminates IO idle time atop vLLM.
    We tried to compare its original open-sourced version,
    but found its performance is {1.2--5.1\,$\times$} slower in TTFT 
    and {1.2--1.9\,$\times$} in TPOT than the chosen vLLM release
    even without memory overloading.
    This is because it was implemented on an old version of vLLM (v0.2.0),
    where important optimizations 
    (\eg{
        FlashAttention/FlashInfer kernels~\cite{flashattention2,flashinfer}, 
        chunked prefill~\cite{sarathi}}) are missing.
    Therefore, we integrated our scheduler and attention backend into the original InferCept 
    for a fair comparison.

    \item \textbf{Llumnix~\cite{llumnix}.} 
    Llumnix adopts load balancing to cope with memory overloading of an instance,
    and migrates KVCache between instances to free sufficient memory in case of
    insufficient memory even with load balancing.
    We compared with the latest version of Llumnix (release v0.1.0).

\end{itemize}

\subsection{End-to-end Results}
\label{sec:eval-e2e}

\begin{figure*}[!t]
        \begin{minipage}{1\linewidth}
        \hspace{-1mm}
        \centering    
        \includegraphics[width=\columnwidth]{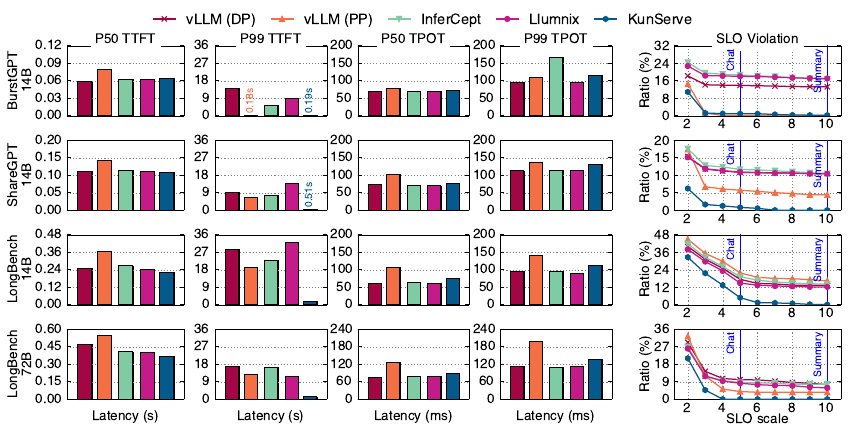} \\[1pt]    
        \end{minipage} \\[-5pt]
        \begin{minipage}{1\linewidth}
        \caption{\small{%
        {
            The end-to-end latency results.
            Column from 1 to 4 is the end-to-end metrics of different workloads.
            The last column is the SLO violation of TTFT and TPOT with different SLO scales.
        }}}
        \label{fig:e2e-latency}
        \end{minipage} \\[-0pt]
\end{figure*}

\nospacestitle{End-to-end serving performance. \,}  
We first measure the end-to-end latency of serving requests 
when running BurstGPT with different datasets on different systems,
where the latency is measured from the client's perspective, 
i.e., the time from the client sending a request to receiving the tokens. 

The second column of {\fig{fig:e2e}} presents how the mean
TTFT changes over time given a measured time window (e.g., {100s}),
and {\fig{fig:e2e-latency}} presents the zoomed-in view of the
P50 and P99 latencies when evaluating different workloads on different models.
First, {\sys} has {12.7--72.2}\,$\times$ faster P99 TTFT than other baselines,
because it frees up sufficient memory under memory overloading,
which enables requests queued in other systems to be served with a larger batch size.
For other systems, they either suffer from recomputation overhead (vLLM),
or queuing overhead waiting for swapping (InferCept) or migration (Llumnix)
under memory overloading.
Specifically, the timeline plotted in \fig{fig:e2e} clearly
shows that the TTFT increase coincides with the increased KVCache demands (the first column in {\fig{fig:e2e}}).

Although vLLM (PP) has a larger KVCache capacity,
it still suffers from medium and tail latency increases
due to the lower throughput.
As shown in the third column of {\fig{fig:e2e}},
the average throughput of PP is {3.3--21.8}\% slower than other systems,
because PP has bubbles during execution. 
Such a lower throughput leads to more KVCache capacity being required under bursts since
pending requests are not digested by the system.
Meanwhile, unlike {\sys} that schedules pending requests to eliminate bubbles,
vanilla pipelined execution cannot simply adopt lookahead batch formulation 
techniques (\textsection{\ref{sec:online-sched}})
because it requires waiting for sufficient requests to be scheduled.
Such waiting also leads to increased end-to-end latencies.

Compared to other baselines, 
{\sys} trades a little increase in P50 TPOT,
and P99 TPOT because it executes requests in a larger batch
to eliminate queuing.
For example, in {LongBench-14B workload}, the P50 TPOT of {\sys} is
{15.8--22.7}\% higher than other baselines.
We believe it is a reasonable trade-off because such increases
are still within the SLOs of targeted applications,
which we describe next.
Interestingly, {\sys} even has a little P50 TTFT improvement 
in the LongBench workload.
This is because the long and {diverse} input of requests in this workload
makes the system more prone to memory overloading caused by severe memory fragmentation~\cite{llumnix}. 
Thus, the many queued requests affect normal requests.

\stitle{SLO attainment.\,}
SLO is an important metric for serving systems~\cite{DBLP:conf/osdi/LiZZL00HCZGS23,distserve},
which defines the maximum acceptable latency for a request.
Requests whose latency exceeds the SLO are not useful because users
may abandon them~\cite{DBLP:journals/corr/abs-2407-00079}.
Because different applications have different maximum acceptable latency requirements (SLOs),
we evaluate the SLO violation of all systems under different SLO scale factors,
similar to previous works~\cite{DBLP:conf/osdi/LiZZL00HCZGS23,splitwise,dynamollm,distserve}.

Specifically,
the last column of {\fig{fig:e2e-latency}} shows the SLO violation of all systems
with different SLO scale factors,
where a scale factor of $N$ means that the maximal tolerable
latency is $N$ times the {P50} latency of the best baseline.
To help understand how the reduced SLO violations of {\sys}
benefit end-to-end applications,
we also mark the typical scale for our evaluated applications,
i.e., we set 5 for chat---a tight SLO as it requires quick
responsiveness, while
for document summarization, we set a looser factor of 10,
following previous works~\cite{distserve}.
We can see that {\sys} achieves {7.2---12.8}\% average SLO violation 
reductions on various workloads,
and more importantly, it almost eliminates all violations
with a scale larger than 4 for all workloads.
Other baselines cannot eliminate SLO violations even with an extremely loose factor of 10
because during bursts, there
are considerable numbers of queued requests suffering from {45---840}\,$\times$ 
tail latency increases.

\stitle{Multi-GPU instance performance.  \,}
Due to space limitations,
we only present the results of the model (Qwen-2.5-72B) that requires multi-GPU
for serving on the LongBench dataset.
Results on other datasets are similar.
As shown in \fig{fig:e2e} and \fig{fig:e2e-latency},
the trend is similar to that of single-GPU instances:
{\sys} reduces the P99 latency by
{8.4--11.9}\,$\times$ compared to other baselines,
at the cost of a slight ({18.3--22.7\%}) increase in P50 TPOT and P99 TPOT.
The multi-GPU model achieves similar results because
each instance (containing multiple GPUs) can be viewed as a whole
as a single logical GPU.
The multi-GPU model benefits even more from dropping parameters
because the relative ratio of parameter memory is large,
as shown in Table~\ref{tab:model}.

\subsection{Ablation Studies}
\label{sec:eval-ablation}

\begin{figure}[!t]
        \begin{minipage}{1\linewidth}
        \hspace{-1mm}
        \centering    
        \includegraphics[width=\columnwidth]{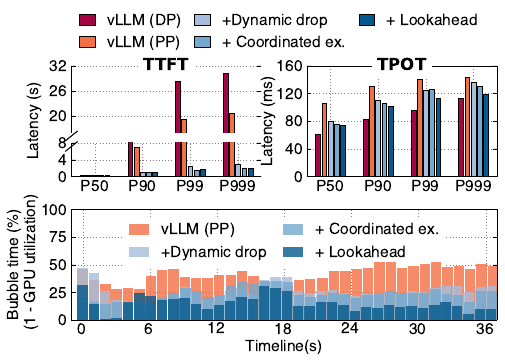} \\[1pt]    
        \end{minipage} \\[-5pt]
        \begin{minipage}{1\linewidth}
        \caption{\small{%
        {
            An ablation study of running {\sys} on {Qwen-2.5-14B} on LongBench dataset.
            A smaller bubble time directly implies a better GPU utilization.
        }}}
        \label{fig:ablation-net}
        \end{minipage} \\[-5pt]
\end{figure}

\noindent
To study the effectiveness of each technique proposed in \textsection{\ref{sec:design}},
we conducted an ablation study on the system performance with different techniques 
incrementally enabled. 
{\fig{fig:ablation-net}} shows the detailed study results on the LongBench dataset
with {Qwen-2.5-14B} model. 
We omit other workloads and models due to space limitation since they have similar results.
We report the end-to-end request latencies 
during the burst period in \fig{fig:e2e}. 

\stitle{Effectiveness of dynamic parameter drop. \,}
First, we can see that parameter drop contributes (+Dynamic drop) to the most
tail latency reductions.
On the LongBench workload,
the P90, P99 and P999 TTFT of {\sys} are reduced by
{8.8\,$\times$, 11.7\,$\times$ and 10.3\,$\times$} compared to vLLM (DP).
The key reason is that it completely eliminates queuing delays.
Specifically, under bursts,
there are {87} queued requests (whose TTFT > SLO(5\,$\times$)) in this evaluation,
{\sys} executes them with enlarged GPU
memory freed by dropping parameters.
Though a larger batch size and pipeline bubbles lead to a TPOT increase 
in request processing ({21--31.9\%} increase compared to the original DP scheduling), 
it is still orders of magnitude smaller than queuing introduced by insufficient
memory of vanilla vLLM.

\stitle{Effectiveness of coordinated exchange. \,} 
Second, with coordinated exchange (+ Coordinated ex.),
{\sys} further reduces the P99 and P999 TTFT by
{1.5\,$\times$, and 1.4\,$\times$} respectively.
Meanwhile, it reduces the P90 and P999 TPOT by
{5\%}.
Coordinated exchange benefits both the TTFT and TPOT because without it,
the prefill of new requests as well as their decode requests
cannot execute smoothly,
because the intermediate activation suffers significant stalls
due to exchanging the KVCache.
Since the exchange time ({1.3s}) is larger than the typical execution time 
(\eg{221ms for prefill and 60ms for decode}), the stall is non-trivial.

\stitle{Effectiveness of lookahead batch formulation. \,}
With lookahead batch formulation (+ Lookahead),
we further reduce the P90, P99, and P999 TPOT by
{4.5\%, 10.6\%, and 9.7\%}, respectively.
The reduction in latency directly comes from the more efficient
pipeline execution:
without lookahead batch formulation,
{\sys} suffers {21.9\%} bubble time (the ratio of idle GPU cycles)
on average during pipelined execution,
while with it, the bubble time is only {8.3\%}.
The reduced bubble time further improves throughput by 20\%.

\begin{figure}[!t]
        \begin{minipage}{1\linewidth}
        \hspace{-1mm}
        \centering    
        \includegraphics[width=\columnwidth]{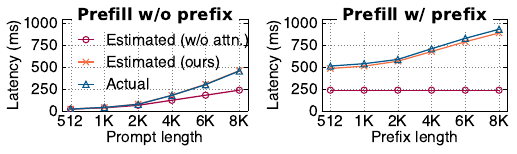} \\[1pt]    
        \end{minipage} \\[-5pt]
        \begin{minipage}{1\linewidth}
        \caption{\small{%
        {
            A comparison of execution latency estimated with our cost model and the 
            real execution time of a Qwen-2.5-14B model in A800 GPUs. 
            Left: the execution without prefix attention while right: the execution with prefix attention.  
        }}}
        \label{fig:cost-model}
        \end{minipage} \\[-5pt]
\end{figure}

\subsection{Accuracy of the batch formulation cost model}
\label{sec:eval-accuracy}

\noindent
To evaluate the accuracy of {\sys}'s cost model described in \textsection{\ref{sec:online-sched}},
we compare it with a baseline cost model neglecting attention computation cost found in existing work~\cite{nanoflow}
and the ground truth.
To demonstrate the generality of our model in both prefill and chunked prefill,
we evaluate both requests without attention chunk ($R4^0$ in {\fig{fig:pp-imbalance}} (c))
and with it ($R4^1$).
As shown in \fig{fig:cost-model},
for both cases,
our cost model shows less than 5\% deviation while
the current formulation without considering attention
has up to 48\% and 74\% deviations for requests without and with prefix attention, respectively.
This confirms the importance of considering the attention computation cost 
in the cost model.

\subsection{Effectiveness of dynamic restoration}
\label{sec:eval-restore}

\begin{figure}[!t]
        \begin{minipage}{1\linewidth}
        \hspace{-1mm}
        \centering    
        \includegraphics[width=\columnwidth]{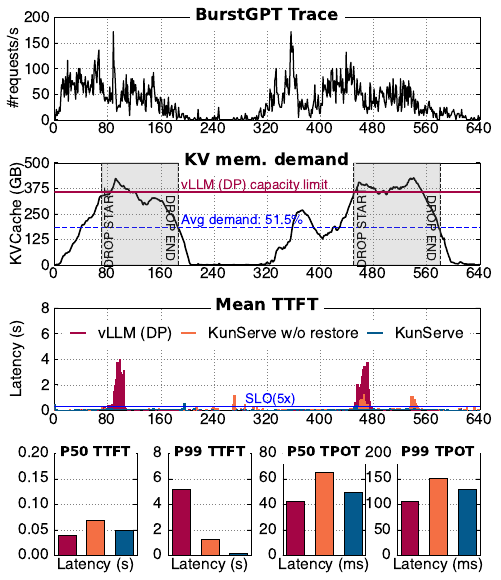} \\[1pt]    
        \end{minipage} \\[-5pt]
        \begin{minipage}{1\linewidth}
        \caption{\small{%
        {
            The performance of {\sys} and its baselines in a 
            long run (640s) of BurstGPT.
        }}}
        \label{fig:long-trace}
        \end{minipage} \\[-5pt]
\end{figure}

\noindent
To show the effectiveness of dynamic parameter restoration,
{\fig{fig:long-trace}} presents the serving performance over a
long run of BurstGPT workload with multiple overloading periods.
To help understand the behavior of {\sys},
we mark the time periods with dropping as grey boxes,
other periods are running without parameter drop. 

First, we observe that dynamic parameter restoration
reduces the P50 latencies of TTFT and TPOT
by 28\,\% and 23\,\%, respectively,
due to the reduction of unnecessary pipeline execution.
Second, restoration improves the P99 TTFT and TPOT
by {6.4\,$\times$} and {1.2\,$\times$}, respectively.
Without restoration,
{\sys} falls back to vLLM (PP), resulting in lower throughput
during normal periods, and consequently suffers from larger bursts with insufficient memory
even with the dropped parameters,
as illustrated at the beginning of the second wave in
the third row of \fig{fig:long-trace} (time 440s).

\subsection{Performance under extreme bursts}
\label{sec:eval-burst}

\begin{figure}[!t]
        \begin{minipage}{1\linewidth}
        \hspace{-1mm}
        \centering    
        \includegraphics[width=\columnwidth]{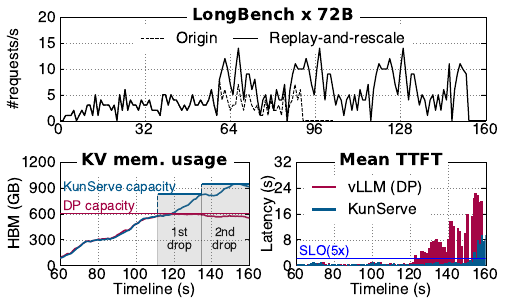} \\[1pt]    
        \end{minipage} \\[-5pt]
        \begin{minipage}{1\linewidth}
        \caption{\small{%
        {
    An evaluation of {\sys} running {Qwen-2.5-72B} under
    extreme bursts.
        }}}
        \label{fig:peak-wave}
        \end{minipage} \\[-5pt]
\end{figure}

\noindent
While {\sys} drops parameters to mitigate queuing,
the memory that can be freed is bounded by the model size (see Table~\ref{tab:model}),
so we have a limit in handling overloading caused by bursts.
Nevertheless, {\sys} can handle bursts much longer than existing systems,
i.e., longer than any burst we have seen in the BurstGPT trace.

To evaluate the limit in handling bursts with {\sys},
{\fig{fig:peak-wave}} shows the performance of {\sys} and vLLM
when running under an unrealistic extreme burst.
Specifically, to evaluate an extreme burst,
we use a BurstGPT setup as follows:
upon meeting the first burst, we repeatedly replay the bursts
until all evaluating systems are out of memory.
The setup is shown in the first row of \fig{fig:peak-wave}
while the second row compares the performance of {\sys} and vLLM (DP).
The evaluated model is Qwen-2.5-72B.
First, {\sys} reaches the memory limit in {152s},
which is {1.5\,$\times$} longer (starting from 60s) than vLLM thanks to the dropped memory.
During this period, {\sys} triggers {2} times of parameter dropping,
resulting in {57\%} incrementally freed KVCache memory.
Before {\sys} reaches the memory limit,
{\sys} meets no SLO (5\,$\times$) violations while vLLM suffers 
up to {42\,$\times$} TTFT increase.

While {\sys} also suffers from latency increases when out of memory,
we don't encounter such a situation under real-world traces.
More importantly,
the much longer standing time of {\sys} allows the serving systems to
smoothly scale up new instances to handle the bursts.

\section{Discussion} 
\label{sec:dislim}

\nospacestitle{Supporting MoE models. \,} 
Our current implementation focuses on dense models, 
while 
Mixture of Experts (MoE) models are becoming increasingly popular recently: 
A key feature is that the inference of a request only activates a small subset of model parameters.
For common MoE serving configurations like expert parallelism (EP)~\cite{DBLP:journals/corr/abs-2412-19437},
{\sys} seamlessly supports them because EP only changes the memory layout within an instance, 
while {\sys} focuses on managing the GPU memory across serving instances. 
More importantly, {\sys} is still effective for MoE models because {\sys} 
only relies on the assumption that the model weights occupy a large portion of the GPU memory on an instance,
which holds even with a sparse activation of experts (see Table~\ref{tab:model}). 
The assumption holds because though a request only requires a small portion of the model parameters,
an instance still needs to load all the (large) model parameters to handle batches of requests 
that may activate all the experts.

\stitle{Compatibility with different parallelism. \,}
{\sys} is compatible with different parallelism in LLM serving.
{\sys} only changes the parameter layout across instances 
in layer granularity,
which is orthogonal to both the intra-layer layout change (e.g., EP and TP) within one instance 
and instance cooperation in SP. 
Thanks to the LLM's modular structure, the intra-instance and inter-instance parallelism techniques can be applied together~\cite{DBLP:journals/corr/abs-1909-08053,loongserve}.

\stitle{Comparison with autoscaling. \,}
Autoscaling---adding more instances to handle overloads---is also a common approach to handle memory overloading~\cite{DBLP:conf/osdi/ZhangWLWS0025}.
A key difference is that {\sys} does not have the cold start time---the time to make an instance capable of serving.
Thus, {\sys} is better than autoscaling in cases where dropping alone is able to handle the overloading,
as the cold start time is typically non-trivial for LLM providers~\cite{DBLP:journals/corr/abs-2407-00079}. 
Nevertheless, for long bursts (see \textsection{\ref{sec:eval-burst}}),
{\sys} still incorporates autoscaling since the memory that can be freed by dropping is limited:
the continuously coming requests from the burst will exhaust all the free memory freed by dropping.

\section{{Related Work}}
\label{sec:related}

\nospacestitle{Handling memory overloading with lossy methods. }
One possible way to handle memory overloading is to reduce the memory footprint of the serving,
e.g., by compressing the activations~\cite{llm-quant,quant-eval}.
For example, {FP8 quantization}~\cite{llm-compressor} reduces the token memory usage by {2}\,$\times$,
and methods like SparseGPT~\cite{sparsegpt} prune parameters to 50\% sparsity.
Unfortunately, such methods are lossy and can lead to model accuracy degradation
or compromised user experience~\cite{quant-damage}. 
{\sys} copes with the performance degradation caused by memory overloading without sacrificing the model accuracy.

\stitle{Handling memory overloading with lossless methods.}
{\sys} continues the line of work on handling memory overloading during LLM serving without modifying the model inference~\cite{vllm,DBLP:journals/corr/abs-2404-09526,DBLP:journals/corr/abs-2407-00079,llumnix,distserve,vllm,infercept}.
These works focus on allowing queued requests to execute by reorganizing GPU memory either
with swap or migration-based methods,
which do not create more space for execution so they either sacrifice ongoing requests or
queued requests, as analyzed in \textsection{\ref{sec:state-of-the-art}}.
In contrast, {\sys} frees more memory for execution with a new parameter-centric memory management method.

\stitle{LLM serving optimizations.}
Considerable research has focused on improving the efficiency of LLM serving
under abundant memory~\cite{flashattention,flashattention2,sarathi-v1,fastgen,vllm,DBLP:journals/corr/abs-2405-04437,distserve,splitwise}.
{\sys} builds on these works and seamlessly integrates with them.
A recent work---{POD-ATTENTION~\cite{podattn}}---proposes a better chunked prefill implementation.
{It is orthogonal to our work and
{\sys} can benefit from its high-performance kernel to get better performance in all states.
NanoFlow~\cite{nanoflow} provides us with a more efficient microbatch scheduling, 
which is of help to {\sys} after parameter dropping.}

\stitle{OS techniques for handling memory overloading.}
Handling memory overloading has been studied in operating systems for decades:
e.g., Linux adopted a swap-based mechanism to handle memory pressure~\cite{mglru}.
{\sys} leverages the domain-specific knowledge of LLM serving to
expose more memory to serving requests beyond the limit of a general-purpose swap-based method.

\section{Conclusion}
\label{sec:concl}

\noindent
In this paper, we are the first to demonstrate that parameter-centric memory management
can effectively address the latency spikes caused by memory overloading in LLM serving.
We built {\sys}, an LLM serving system that cooperatively drops parameters
to free up memory to
eliminate queuing under overloading.
We also proposed a set of techniques to ensure all requests execute efficiently
after parameter dropping,
including drop plan generation with local unified memory management,
coordinated KVCache exchange and lookahead batch formulation.
{\sys} reduces tail TTFT by up to {72.2}\,$\times$ compared to state-of-the-art systems
like Llumnix, vLLM and InferCept.

\section{{Acknowledgement}}
\label{sec:ack}

\noindent
We sincerely thank our shepherd Jayashree Mohan and
the reviewers from OSDI'25 and EuroSys'26 for
their insightful feedback.
We are grateful to
Wencong Xiao from ByteDance,
Mingcong Han, Hanze Zhang, Xian Xu, Yu Xia, Yingyi Hao, and Hongrui Xie from IPADS for
their valuable advice. We also thank the ByteDance seed-foundation team
for their platform support. 
We thank Chao Fei from KAUST for his contributions to the codebase of {\sys}.
This work was supported in part by
the National Natural Science Foundation of China (No. 62572302 and 62272291),
and the Fundamental Research Funds for the Central Universities. 

\balance

\small{
\bibliographystyle{acm}
\bibliography{balloon}
}

\clearpage



\end{document}